\documentclass[doublecol]{epl2} 
\usepackage{psfrag}

\title{Exact Haldane mapping for all $S$ and super universality in spin chains}

\author{ A.M.M.~Pruisken\inst{1}, R. Shankar\inst{2} and 
N. Surendran\inst{1,3}}
\shortauthor{Pruisken \etal}

\institute{
\inst{1}Institute for Theoretical Physics, Valckenierstraat 65,
  1018 XE Amsterdam, The Netherlands\\
\inst{2}The Institute of Mathematical Sciences, CIT Campus,
  Chennai 600 113, India.\\
\inst{3}Present address: ICTP, Strada Costiera 11, 34014 Trieste, Italy.
}

\abstract{
The low energy dynamics of the anti-ferromagnetic Heisenberg spin
$S$ chain in the semiclassical limit $S\rightarrow\infty$ is known
to map onto the $O(3)$ nonlinear $\sigma$ model with a $\theta$
term in $1+1$ dimension. Guided by the underlying dual symmetry of
the spin chain, as well as the recently established topological
significance of ``dangling edge spins," we report an {\em exact}
mapping onto the $O(3)$ model that avoids the conventional large
$S$ approximation altogether. Our new methodology demonstrates all
the super universal features of the $\theta$ angle concept that
previously arose in the theory of the quantum Hall effect. It
explains why Haldane's original ideas remarkably yield the correct
answer in spite of the fundamental complications that generally
exist in the idea of semiclassical expansions.
}

\begin{document}

\maketitle

In $1983$, Haldane proposed that the low energy
dynamics of the anti-ferromagnetic Heisenberg spin $S$ chain can
be taken from the $O(3)$ nonlinear $\sigma$-model (NLSM) in $1+1$
dimension and in the presence of the $\theta$ term.~\cite{Haldane}
At least within the limitations of a large $S$ approximation the
parameter $\theta$ was found to take on the values $0$ or $\pi$
only, dependent on whether $S$ is integral or half-integral
respectively. Since the standard $O(3)$ model with $\theta=0$ is
known to display a mass-gap,~\cite{BrezinZinn-Justin} Haldane
concluded that the integral spin chain is always gapped. This is
unlike the $S=1/2$ chain, for example, which from the Bethe ansatz
solution is known to display gapless excitations ~\cite{Bethe}.
Based on numerical work it is now generally accepted that the
uniform integral spin chain is always gapped whereas the half
integral spin chain is generally gapless .

It has recently been pointed out, however, that a universal
topological feature of the $\theta$ angle concept has historically
been overlooked.~\cite{PruiskenShankarSurendran} The $\theta$
vacuum quite generally displays {\em massless chiral edge
excitations}~\cite{PruiskenSkoricBaranov} that have important
consequences for the theory on the strong coupling side. Within
the grassmannian $U(M+N)/U(M) \times U(N)$ nonlinear sigma model
approach to localization and interaction phenomena, for example,
it was shown that the massless edge excitations are directly
related to the existence of {\em robust topological quantum
numbers} that explain the stability and precision of the quantum
Hall effect.~\cite{PruiskenBurmistrov} This has led to the idea of
{\em super universality} of quantum Hall physics that, unlike the
common belief in the field, is independent of the details of the
theory such as the number of field components or, for that matter,
replica limit $M,N \rightarrow 0$.~\cite{PruiskenBurmistrov}

Based on the Haldane mapping it is next natural to expect that the
super universality statement can be extended to also include
quantum spin liquids. Indeed, renormalization group studies have
clearly indicated that the dimerised $S=1/2$ spin chain displays
all the basic features of the quantum Hall
effect~\cite{PruiskenShankarSurendran} in much the same manner as
what has recently been observed,for example, in the exactly
solvable large $N$ limit of the $CP^{N-1}$
model.~\cite{PruiskenBurmistrovShankar} The dynamics of the {\em
dangling edge spins} of the chain thereby plays a role that is in
many ways the same as that of the {\em massless chiral edge
excitations} of the $\theta$ vacuum. Unfortunately, a generalized
Haldane mapping that includes the dangling spins at the edges of
the spin $S$ chain has sofar been obtained only for the
semiclassical $S=\infty$ limit.~\cite{PruiskenShankarSurendran}
Given the extensive literature on the subject of quantum spin
chains, it is somewhat surprising to know that not a single
attempt has been reported that would in principle resolve this
longstanding drawback and extend the Haldane mapping to include
finite values of $S$.

One of the main objectives of the present Letter is to show that
the statement of super universality of quantum spin liquids cannot
in general be established by using the semiclassical large $S$
idea alone. To illustrate the problems we first recall the
hamiltonian of the spin chain
\begin{equation}\label{hamiltonian}
 \mathcal{H} = \frac{J}{S} \sum_I \Big({\bf S}_{I 1} \cdot {\bf
 S}_{I 2} + \kappa ~{\bf S}_{I 2} \cdot {\bf S}_{I\!+\!1~ 2} \Big).
\end{equation}
Here, $J > 0$ favors the anti-ferromagnetic (N\'{e}el) ordering,
the integer $I$ is the dimer index and the subscripts $1$ and $2$
denote the two sites within each dimer. The variable $\kappa$ with
$0<\kappa < \infty$ is the dimensionless nearest neighbor coupling
between the dimers. The path integral representation involves
$O(3)$ vectors $\hat {\bf n} (t)$ with $\hat {\bf n}^2 =1$ and $t$
denoting the imaginary time. The action is ~ \cite{FradkinStone}
\begin{eqnarray}
 \mathcal{S} &=& i S ~\sum_{I}\Omega[{\hat {\bf n}}_{I1}] +
 \Omega[{\hat {\bf n}}_{I2}] + \nonumber \\
 &&  S J \sum_{I} \oint ~\big(
 {\hat {\bf n}}_{I 1} \cdot {\hat {\bf n}}_{I 2} + \kappa~{\hat {\bf n}}_{I 2}\cdot
 {\hat {\bf n}}_{I+1~ 1} \big) \label{part-fn}
\end{eqnarray}
The quantity $\Omega[{\hat {\bf n}}_{I \alpha}]$ is the solid
angle term associated with each lattice site $I \alpha$ and $\oint
= \int_0^\beta dt$. To extract the low energy dynamics of the spin
chain from Eq. \ref{part-fn} several basic assumptions are
necessary. The historical procedure was based on the change of
variables~ \cite{Affleck}
\begin{equation}
\label{mltrans}
 {{\bf m}}_I = \frac{1}{2} {(\hat{{\bf n}}_{I1} - \hat{{\bf n}}_{I2})},~~
 {\bf l}_I = \frac{1}{2} {(\hat{{\bf n}}_{I1} + {\hat{\bf n}}_{I2})}
\end{equation}
where $\hat{{\bf m}}_I = {\bf m}_I /|{\bf m}_I|$, which is a
measure of the N\'{e}el ordering, is taken as the {\em soft mode}
in the problem that should be retained. The field variable ${\bf
l}_I$, on the other hand, is taken as the {\em hard mode} that
should be integrated out. The simplest way to do this is by using
semiclassical approximations. Assuming $S\rightarrow\infty$ then
Eq. \ref{part-fn} can be evaluated at the saddle point which in
the long wavelength limit (slowly varying $\hat{{\bf m}}_I$)
yields the NLSM in the presence of the $\theta$ angle
~\cite{Affleck,PruiskenShankarSurendran}. The complications,
however, occur in the computation of the $1/S$ corrections that
all diverge as $\beta \rightarrow \infty$. The origin of the
divergencies are
easily understood from the following {\em exact} expression
for the solid angle term of each dimer.~\cite{Surendran}
\begin{equation}
 \label{solid-angle}
 \Omega [ {\hat {\bf n}}_{I1} ] + \Omega [ {\hat {\bf n}}_{I2} ]
 = 2~ \oint {\bf l}_I \cdot {\hat {\bf m}}_{I} \times
 \partial_t  {\hat {\bf m}}_{I} .
\end{equation}
The action of the ``hard" modes has therefore no time derivatives
and this, in turn, implies that {\em coincident} operators ${\bf
l}_I (t)$ have a divergent expectation value. What this simple
example is telling us is that semiclassical expansions are quite
generally plagued by ambiguities that are inherent to the bosonic
path integral representation of Eq. \ref{part-fn}. These
ambiguities, as we shall see later on in this Letter, are the main
reason why the general theory of the Heisenberg anti-ferromagnet
has not advanced beyond the naive $S=\infty$ saddle point limit.

\begin{figure}
\psfrag{1}[c,t][c,t]{$1$}
\psfrag{2}[c,t][c,t]{$2$}
\psfrag{3}[c,t][c,t]{$3$}
\psfrag{one}[l,b][l,b]{$1$}
\psfrag{k}[c,b][c,b]{$\kappa$}
\begin{center}
\includegraphics[width = .25 \textwidth]{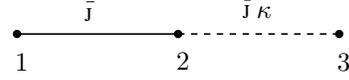}
\caption{The $3$-spin problem with relative bond strengths $1$ and
$\kappa$.} \label{3spins}
\end{center}
\end{figure}

\noindent{\bf$\bullet$ {\em Three spin problem.}~}As the most
important step next we introduce a novel mapping onto the NLSM
that avoids the use of semiclassical approximations altogether.
This is accomplished if, instead of Eq. \ref{mltrans}, we pursue
an effective action of the spin chain as defined by {\em
decimation}, i.e. by eliminating the spins on one sublattice (say,
$\hat{{\bf n}}_{I1}$) while retaining those on the other
($\hat{{\bf n}}_{I2}$). The primary focus will be on {\em open}
spin chains containing $2N+1$ sites since they display both the
subtleties of ``dangling edge spins" and ``dual invariance." By
dual invariance we mean that the dimerised spin system described
by Eqs. \ref{hamiltonian} and \ref{part-fn} is invariant under
the interchange between the weak and strong bonds
\begin{equation}\label{dual-trfrm}
J \rightarrow J\kappa,~~~~~~\kappa \rightarrow \frac{1}{\kappa} .
\end{equation}
For simplicity we first consider the three spin system as sketched
in Fig. \ref{3spins} which is the simplest possible system with
edges that is manifestly self-dual.
It is advantageous to express the $O(3)$ vector field $\hat {\bf
n}$ in terms of a grassmannian field variable $Q = {\bf {\hat n}}
\cdot {\bm \tau} \in SU(2)/U(1)$, with $\bm \tau$ denoting the
Pauli matrices. This leads to the three spin action
\begin{eqnarray}
 \mathcal{S} [Q_1 , Q_2 , Q_3] &=& i S \Big \{ \Omega[Q_1]
+ \Omega[Q_2] + \Omega[Q_3] \Big \} + \nonumber \\
\label{Grassmann-3}
 && \frac{1}{2} SJ \oint ~tr \Big(Q_1 + \kappa~ Q_3 \Big) Q_2  .
\end{eqnarray}
We can always write $Q = T^{-1} \tau_z T$ with $T \in SU(2)$ such
that the solid angle term can be expressed explicitly according to
$i\Omega[Q] = i\Omega[T] =\oint tr T \partial_t T^{-1} \tau_z$
~\cite{PruiskenShankarSurendran}. The decimation of the three spin
system is defined as follows
\begin{equation}\label{map-3}
 e^{-{\mathcal{S}}_{eff} [Q_1 , Q_3]} = \int \mathcal{D}[Q_2]
~e^{-{\mathcal{S}} [Q_1 , Q_2 ,
 Q_3]} .
\end{equation}
To evaluate ${\mathcal{S}}_{eff}$ we notice that the matrix $Q_1 +
\kappa Q_3$ in Eq.~\ref{Grassmann-3} is hermitian and traceless.
Therefore write
\begin{eqnarray}\label{Q-U}
 J \left( Q_1 + \kappa Q_3 \right)&=& B V^{-1} \tau_z V \\
 B &=& J(1+\kappa) \sqrt{ 1- \frac{\kappa}{2(1+\kappa)^2}
 tr \delta Q \delta Q } ~~~~~\label{bmod-1}
\end{eqnarray}
where $V \in SU(2)$ and $\delta Q = Q_3 - Q_1$ which is taken as a
small quantity. Similarly, we expand the matrix $V$ in powers
of the differential $\delta T = T_3 - T_1$
\begin{equation}\label{U-U}
 V = T_1 + \frac{\kappa}{1+\kappa} \delta T + \mathcal{O}(\delta^2).
\end{equation}
Based on Eqs \ref{Q-U} - \ref{U-U} we obtain
${\mathcal{S}}_{eff}$ order by order in a derivative expansion. To
see this we replace $Q_2\rightarrow V^{-1}Q_2 V$ such that Eq.
\ref{Grassmann-3} can be rewritten as
\begin{eqnarray}
 \mathcal{S} = i S \Big ( \Omega[T_1] &+& \Omega[T_3] \Big ) +
 \mathcal{S}_0 [Q_2] + \tilde{\mathcal{S}} [Q_1 , Q_2 ,
 Q_3] ~~~\label{Grassmann-3a} \\
\label{One-spin-in-B}
 \mathcal{S}_0 [Q_2] &=& S\oint tr \left( T_2 \partial_t T_2^{-1}
 +\frac{B_0}{2}  Q_2 \right) \tau_z ~~~~\\ \label{Grassmann-3b}
 \tilde{\mathcal{S}} [Q_1 , Q_2 ,
 Q_3] &=&  S \oint tr \left( V \partial_t V^{-1} + \frac{\delta
 B}{2} \tau_z \right) Q_2
\end{eqnarray}
with $B_0 = J(1+\kappa)$ and $\delta B = B - B_0$. Notice that the
piece $\mathcal{S}_0$ in Eqs \ref{Grassmann-3a} and
\ref{One-spin-in-B} is the action of a single spin $S$ in a
constant magnetic field $B_0$ which is {\em exactly} solvable. Of
interest are the non-vanishing one and two point correlations of
the $Q_2 = Q_2^{\sigma\sigma^\prime}$ matrix field~
\cite{PruiskenSkoricBaranov,PruiskenShankarSurendran}
\begin{eqnarray}\label{cum-exact-1}
 \langle Q_2 \rangle &=& - \tau_z ,~~~
 \langle Q_2^{21} (0) Q_2^{12} (t) \rangle = \frac{2}{S}
 \vartheta (t) e^{- B_0 t} .
\end{eqnarray}
Here, the limit $\beta\rightarrow\infty$ is understood and
$\vartheta (t)$ denotes the Heaviside step function. The piece
$\tilde{\mathcal{S}}$ in Eqs \ref{Grassmann-3a} and
\ref{Grassmann-3b} contains derivatives acting on the ``soft"
modes $Q_1$ and $Q_3$ and can be treated in a cumulant expansion.
This leads to ${{\mathcal{S}}_{eff}}$ obtained to lowest orders in
$\partial_t$ and $\delta$
\begin{eqnarray}\label{Seff-exact1a}
&&\! \!\!{{\mathcal{S}}_{eff}} [Q_1, Q_3] = i S \Omega[T_3]
+~ S  \frac{\kappa }{\kappa +1} \times \nonumber \\
&&\times \oint tr \left\{ \frac{J}{4} (\delta Q)^2
 + \frac{1}{J\kappa} (\partial_t Q)^2  - \delta ( T \partial_t
 T^{-1} \tau_z)\right\}\!.~~~~~
\end{eqnarray}

\noindent{\bf $\bullet$ {\em Exact mapping.}~} Eq.
\ref{Seff-exact1a} can directly be generalized to obtain the
effective action ${\mathcal{S}}_{exact}^{o}$ for the dimerised
spin chain with $2N+1$ sites and the result is
\begin{eqnarray}\label{Seff-exact2}
{\mathcal{S}}_{exact}^{o} &=& \sum_{m=1}^{N} \Big \{
\mathcal{S}_{eff} [Q_{2m-1}
,Q_{2m+1}]~-~ iS~  \Omega[T_{2m+1}] \Big \} ~~~ \nonumber \\
&& ~~+ i S~\Omega[T_{2N+1}] .
\end{eqnarray}
In the continuum limit we obtain the NLSM
\begin{eqnarray} \label{full-action}
 {\mathcal{S}}_{exact}^{o} &=& \frac{1}{g} \int tr \Big \{ \frac{1}{c} ( \partial_t Q )^2
 + c ~ ( \partial_x Q )^2 \Big \} + i\theta ~\mathcal{C} [Q]  \nonumber \\
 &&~~ + i S~ \Omega[T(L)] .
\end{eqnarray}
Here, $\int$ denotes the space-time integral $\oint dt \int_0^L
dx$, $L=2Na$ with $a$ the lattice constant. $\mathcal{C} [Q]$ is
the {\em topological charge} of the matrix field $Q$
\begin{equation}
 \mathcal{C} [Q] = \frac{1}{16\pi i} \int tr \epsilon_{\mu\nu}
 Q\partial_\mu Q \partial_\nu Q = \frac{\Omega [ T(0)
 ] - \Omega [ T(L) ]}{4\pi} ~~~~\nonumber
\end{equation}
and the NLSM parameters are given by~\cite{Sinfty}
\begin{eqnarray}\label{nlsmpar}
 \theta (\kappa)= 4 \pi S \frac{\kappa}{1+\kappa},~{g} (\kappa)=
 \frac{1+\kappa}{S \sqrt\kappa}, ~
 c (J, \kappa)= a J \sqrt{\kappa} .~~
\end{eqnarray}
We will next employ these results to draw important conclusions
that are valid for arbitrary values of $S$.

\noindent{\bf $\bullet$ {\em Duality and super universality.}~}
The solid angle term in Eq. \ref{full-action} is clearly the
action of a single ``edge spin" that makes all the difference
between an {\em open} spin chain and {\em closed} one. In
anticipation of the fact that the ``bulk" of the system is
periodic in the ``angle" $\theta$ we write
\begin{equation}
 \theta(\kappa)=\theta_B (\kappa)+ 2\pi k(\kappa)
 \label{theta-split}
\end{equation}
where $-\pi < \theta_B (\kappa) \leq \pi$ denotes the {\em
fractional} piece and $k(\kappa) = 0, 1, \dots
2S$ the {\em integral} piece of $\theta(\kappa)$. Notice that under the transformation of
Eq.\ref{dual-trfrm} the NLSM parameters
are replaced by
\begin{eqnarray}
 \theta_B (1/\kappa) &=& -\theta_B (\kappa), ~~~~ k(1/\kappa)=S - k(\kappa) \nonumber\\
 g(1/\kappa) &=& g(\kappa),~~~ c(J\kappa, 1/\kappa)=c(J, \kappa).
\end{eqnarray}
Eq. \ref{theta-split} therefore permits a splitting of Eq.
\ref{full-action} into a ``bulk" part $\mathcal{S}_{B} [Q]$ and
an ``edge" part $\mathcal{S}_{E}^{o} [T]$ that are decoupled under
the dual transformation. Specifically, we rewrite
\begin{eqnarray}\label{Seff-exact3}
 {\mathcal{S}}_{exact}^{o} &=& \mathcal{S}_{B} [Q]
+ \mathcal{S}_{E}^{o} [T]\\ \label{Seff-exact3a}
 \mathcal{S}_{B} [Q] &=& \frac{1}{g} \int tr \left\{ \frac{1}{c} (\partial_t Q)^2 + c~ (
\partial_x Q )^2 \right\} + i {\theta_B} \mathcal{C} [Q] ~~~~~\\\label{Sedge-odd}
 \mathcal{S}_{E}^{o} [T] &=& i \frac{k}{2}~\Omega[T (0)]
 +i(S - \frac{k}{2})~\Omega[T (L)] .
\end{eqnarray}
These final expressions are the principal results of this Letter.
Notice that Eq. \ref{Sedge-odd} is a fundamental statement made
on the ``edge" of the $\theta$ vacuum that has traditionally gone
unnoticed.~\cite{PruiskenShankarSurendran,PruiskenBurmistrov}
Following Eqs \ref{One-spin-in-B} and \ref{cum-exact-1} it is
the {\em critical} action of ``dangling edge spins" and should
therefore be regarded as an integral part of the low energy
dynamics of the spin chain. To understand this aspect of the
problem we first consider the class of topological field
configurations $Q_0$ for which $\mathcal{C} [Q_0]$ is strictly an
integer. As is well known, this class sets the stage for the
``instanton picture" of the $\theta$ angle and is geometrically
defined by identifying the edge as a single point, i.e. $Q_0$ is a
constant matrix along the edge of the system, say $Q_0 =
\tau_z$.~\cite{PruiskenBurmistrov} The action for the edge Eq.
\ref{Sedge-odd} is now a trivial phase factor and we immediately
recognize ${\mathcal{S}}_{exact}^{o} = \mathcal{S}_{B} [Q_0]$ as
the theory of the ``bulk" of system that only depends on
$\theta(\kappa)$ modulo $2\pi$.

To incorporate the edge excitations we write $Q= U^{-1} Q_0 U$.
Here $U \in SU(2)$ represents the ``fluctuations" about the
boundary condition $Q_0 =\tau_z$ that carry a {\em fractional}
topological charge. Discarding unimportant phase factors Eq.
\ref{Sedge-odd} now reads $\mathcal{S}_{E}^{o} [T] =
\mathcal{S}_{E}^{o} [U]$ indicating that the matrix $U$ is the
basic field variable for the ``edge." Of primary interest is the
effective action for ``edge" excitations
$\tilde{\mathcal{S}}_{E}^{o} [U]$ which is given by
~\cite{PruiskenBurmistrov,PruiskenShankarSurendran}
\begin{eqnarray}\label{eff-edge}
 e^{-\tilde{\mathcal{S}}_{E}^{o} [U]} =
 e^{-\mathcal{S}_{E}^{o} [U]} \int_{\partial V} \mathcal{D} [Q_0]
 e^{-\mathcal{S}_{B} [U^{-1} Q_0 U]}.
\end{eqnarray}
Here, ${\partial V}$ reminds us of the boundary condition $Q_0 =\tau_z$.
Next, we make use of the fact that the two dimensional theory of
Eq. \ref{Seff-exact3a} develops
a {\em mass gap} when $\theta_B (\kappa) \approx 0$ or $\theta (\kappa) \approx 2\pi k(\nu)$.
~\cite{BrezinZinn-Justin} A finite mass gap in the ``bulk" means that the functional integral
of Eq.\ref{eff-edge} is insensitive to changes in the boundary conditions and, hence,
$\tilde{\mathcal{S}}_{E}^{o} [U] \equiv {\mathcal{S}}_{E}^{o} [U]$ except for $U$
independent terms. We therefore conclude that the dimerised spin chain with varying
values of $\kappa$ displays $2S+1$ different topological phases, labelled by the
integer $k(\kappa)$, where the low energy excitations are solely those of the
``dangling edge spins" with quantum numbers ${k (\kappa)}/{2}$ and $S- {k (\kappa)}/{2}$
respectively. These phases must in general be separated by {\em quantum phase transitions}
(or a vanishing mass gap in the bulk) occurring at intermediate values of
$\kappa$ where $\theta_B (\kappa)$ makes a ``jump" from $+\pi$ to $-\pi$. Notice that
gapless excitations are in general necessary in order for the spin
chain to be able to {\em transport} a spin $\frac{1}{2}$ quantum
over {\em macroscopic} distances from one edge to the other.
Moreover, in complete analogy with the ``electrodynamics picture" of
the $\theta$ angle ~\cite{Coleman} one
may interpret the ``jump" in $\theta_B (\kappa)$
in terms of the creation of ``Coleman charges" that move to the
opposite edges such as to maximally shield the ``background
electric field" $\theta (\kappa)$. ~\cite{misleading}

In summary, emerging from Eqs. \ref{Seff-exact3} -
\ref{eff-edge} are precisely the {\em super universal} features
of the $\theta$ vacuum concept that have previously been
discovered in the context of the quantum Hall effect
~\cite{PruiskenBurmistrov}. The statement of super universality
becomes all the more transparent when, for example, the ``dangling
edge spins" are identified with the phenomenon of {\em massless
chiral edge excitations}, the spin chain parameter
$\theta(\kappa)/2\pi$ is replaced by the {\em filling fraction} of
the Landau levels and, finally, the integer $k(\kappa)$ is
recognized as the {\em robustly quantized} Hall conductance
~\cite{PruiskenShankarSurendran}. For completeness we list the
results for Eq. \ref{Sedge-odd} for {\em open} spin chains
($\mathcal{S}^{e}_{E}$) and {\em closed} ones ($\mathcal{S}^p_E$)
that contain an {\em even} number of sites
\begin{eqnarray}\label{Sedge-exact}
 \mathcal{S}^e_E [U] = i \frac{k}{2}~\Big \{ \Omega[U (0)]
 -\Omega[U (L)] \Big \} , ~~ \mathcal{S}_{E}^{p} [U] &=& 0.~~~
\end{eqnarray}
\noindent{\bf $\bullet$ {\em Semiclassical theory.}~} Next, it is
of interest to know what our explicit and exact results teach us
about the problem of semiclassical expansions that to date has
spanned the subject. For this purpose we address the single spin
problem of Eq. \ref{One-spin-in-B} and employ the Schwinger
boson representation $Q_2^{\sigma\sigma^\prime}=
\delta_{\sigma\sigma^\prime} - 2 z_\sigma z^*_{\sigma^\prime}$
with ${\bf z}^* \cdot {\bf z} = 1$. Fixing the $U(1)$ gauge such
that $z_1=z_1^*=\sqrt{1-z_2^*z_2}$ the $Q_2$ matrix field can be
written as
\begin{eqnarray}\label{matrix-Q}
 Q_2 = - \left(
 \begin{array}{cc}
 1 - 2 z_2^* z_2 & 2 z_2 ~ \sqrt{1-z_2^* z_2} ~~~\\ 
& \\
 ~~~2 z_2^*  ~\sqrt{1-z_2^* z_2} & -1 + 2 z_2^* z_2
 \end{array} \right) .
\end{eqnarray}
The single spin action of Eq. \ref{One-spin-in-B} becomes simply
\begin{equation}\label{harm-osc}
 {\mathcal{S}_0}= 2S \oint \Big \{ z_2^* (t) ~\partial_t~ z_2 (t) +
 B_0~ z_2^* (t-\epsilon) z_2 (t) \Big \} .
\end{equation}
For reasons to be explained shortly, we have introduced an
infinitesimal {\em time splitting} quantity $\epsilon>0$ in the
definition of ${\mathcal{S}_0}$. In the large $S$ limit the
propagator becomes
\begin{equation}\label{propa}
 \langle z^*_2 (0) z_2 (t) \rangle = \frac{1}{2S} \vartheta (t-\epsilon)
 e^{-B_0 t} .
\end{equation}
It can be shown that Eqs. \ref{matrix-Q} - \ref{propa} are
completely equivalent to the Holstein-Primakoff representation of
the single spin. The effect of our introduction of the time
splitting quantity $\epsilon$ in Eq. \ref{harm-osc} and
\ref{propa} is to render the expectation value of {\em
coincident} operators identically equal to zero. The remarkable
conclusion that one can draw from all this is that the exact
correlations of Eq. \ref{cum-exact-1} are obtained in the limit
$S \rightarrow \infty$ and the corrections are zero to all orders
in $1/S$ ~\cite{Klauder}.

The subtle but crucial feature of {\em coincident} operators is
generally lost in Eqs. \ref{part-fn} and \ref{One-spin-in-B},
a drawback of the bosonic path integral that explains why the
traditional saddle point or large $S$ ideas are complicated. On
the other hand, the present theory fundamentally resolves these
ambiguities since our exact and semiclassical results can be
compared directly with the knowledge obtained from different
sources, notably the hamiltonian formalism. Finally, our
methodology is not specifically designed for $SU(2)$ spins alone.
It can also be applied to the $SU(n)$ case, for example, as well
as the Heisenberg anti-ferromagnet in higher dimensions.
\cite{PruiskenShankarSurendran1}

\acknowledgments
This research was funded in part by the Dutch science foundations
FOM and NWO and the EU-Transnational Access program (RITA-CT
2003-506095). One of us (A.M.M.P.) is indebted to the Institute of
Mathematical Sciences (Chennai), Ecole Normale Superieure (Paris)
and the Weizmann Institute (Rehovot) for visiting appointments.


\begin{thebibliography}{0}
\bibitem{Sinfty} As explained later in the text the same expressions are
obtained in the naive $S=\infty$ approximation.
\cite{Affleck,PruiskenShankarSurendran}
\end{thebibliography}
\end{document}